**Dislocation response to electric fields in strontium titanate: A mesoscale indentation study**

Alexander Frisch[1]*, Daniel Isaia[2], Oliver Preuß[2], Xufei Fang[1]*

[1]Institute for Applied Materials, Karlsruhe Institute of Technology, Karlsruhe, Germany

[2]Department of Materials and Earth Sciences, Technical University of Darmstadt, Darmstadt, Germany

*Corresponding authors: alexander.frisch@kit.edu (AF); xufei.fang@kit.edu (XF)

**Abstract:**

Dislocations in perovskite oxides have drawn increasing research interest due to their potential of tuning functional properties of electroceramics. Open questions remain regarding the behavior of dislocations concerning their stability under strong externally applied electric fields. In this study, we investigate the dielectric breakdown strength of nominally undoped $SrTiO_3$ crystals after the introduction of high-density dislocations. The dislocation-rich samples are prepared using the Brinell scratching method, and they consistently exhibit lower dielectric breakdown strength as well as a larger scatter in the breakdown probability. We also study the impact of electric field on the introduction and movement of dislocations in $SrTiO_3$ crystals using Brinell indentation coupled with an electric field of 2 kV/mm. No changes on the dislocation plastic zone size, depth, and dislocation distribution are observed under this electric field. Based on the charge state of the dislocations in $SrTiO_3$ as well as the electrical and thermal conductivity modified by dislocations, we discuss the forces induced by the electric field to act on the dislocations to underline the possible mechanisms for such dislocation behavior.

## 1. Introduction

Dislocations in perovskite oxides have gained increasing research interest due to their proposed impact on functional properties like enhanced large-signal piezoelectric coefficient $d_{33}^*$ in $BaTiO_3$ [1], or increase in the superconducting transition temperature $T_c$ subsequent to plastic deformation in $SrTiO_3$ [2]. It is suggested that functional properties of ceramics are influenced by the elastic stress field around a dislocation, coupled with the charged core of a dislocation surrounded with a space charge layer in ceramics with ionic/covalent bonding [3]. While the elastic stress field of a dislocation is well understood by the long history of research on both metals and ceramics, the charge of a dislocation core as well as the surrounding space charge layer, especially in perovskite oxides, have only recently started to attract more attention [4].

The interplay between dislocations in ceramics and an external electric field was probably first discovered in 1933 from electrical phenomena during the plastic deformation of NaCl, a rock salt structure model material [5]. As later comprehensively reviewed by Whitworth in 1974, mobile dislocations in the rock salt structure are charge-neutral themselves, but can interact with charge-carrying vacancies and thus acquire charges [6]. This results in two main effects. First, an electric field surrounding a charged dislocation which can interact with other external fields, or attract other charged defects forming a space charge layer around the dislocation [6]. Second, during dislocation motion, a charged dislocation can contribute to a separation of charge carriers, resulting in a measurable potential difference [6]. These phenomena are not only observable in the plastically deformable rock salt structure crystals, but also in the plastically deformable II-VI semiconductors of the sphalerite and wurtzite structure, e.g. ZnS or CdTe, as reviewed by Osip'yan et al. in 1986 [7]. In these materials, the mobile dislocations carry partial charges even without interaction with point defects, which was most recently utilized to facilitate the dislocation motion in ZnS through electric fields in transmission electron microscope without the application of mechanical stresses [8].

One of the limiting factors in dislocation-engineering the functional properties of perovskite ceramics has been the introduction of dislocations with high densities into large volumes [9]. With the above-mentioned gain in dislocation mobility under an applied electrical field, the question of whether the mechanically induced dislocation motion in perovskite oxides can be enhanced by an external electric field. If so, it will be desirable to use mechanically seeded dislocations, coupled with external electric field, to achieve more plastically deformable ceramics, namely by way of the so-called *electroplasticty* in ceramics. To this regard, the structure of the dislocation cores, and accordingly also their charge state, has recently been under debate [10], with simulation results showcasing the uncharged nature of most dislocations in $SrTiO_3$, which is a model perovskite oxide [11]. Furthermore, it is remains experimentally elusive whether dislocations in perovskites can interact with vacancies to acquire charges, just as dislocations in NaCl or other oxides such as MgO with rock salt structure. Finally, if dislocations in perovskite oxides can be moved by electric fields, concerns arise about the stability of future devices that use pre-designed dislocation arrays for achieving the functionality.

In this study, we investigate the behavior of pre-engineered dislocations in the perovskite oxide $SrTiO_3$ under external electric fields. To test the stability of dislocation-engineered ceramics, dielectric breakdown experiments were first conducted on samples with a high dislocation density. Furthermore, to investigate the impact of an electric field on dislocation generation and motion, a Brinell indentation setup coupled with an electrical source was designed to conduct plastic deformation assisted by an electric field. The results have insight into the feasibility of electroplasticity in perovskite oxide crystals, as well as the stability of engineered dislocations for possible future functional devices.

## 2. Methods

### 2.1 Material selection

The tested samples were Verneuil-grown, nominally undoped strontium titanate ($SrTiO_3$) single crystals (Alineason Materials Technology GmbH, Frankfurt am Main, Germany). $SrTiO_3$ has a cubic structure at room temperature and was discovered to be plastically deformable in bulk compression at room temperature [12]. Its room-temperature slip systems are of the <110>{110}-type with a yield strength of ~112 - 145 MPa in uniaxial bulk compression along the [001] direction, depending on the pre-existing dislocation densities of $10^{10}$ to $10^{11}$ m$^{-2}$ [13]. For the dielectric breakdown experiments samples of the dimensions 3 x 3 x 0.5 mm$^3$ were used. Brinell indentation tests were conducted on one sample of 10 x 10 x 1 mm$^3$ size. All samples used have one polished (001) surface with no scratches visible under optical microscopy. The pre-existing dislocation density in these samples was checked to be ~$10^{10}$/m$^{-2}$ using the chemical etching method [14].

### 2.2 Dielectric breakdown tests

To account for the possible scattering of the results, 25 samples were used for reference dielectric breakdown measurements, and 17 samples were pre-engineered with high dislocation densities using the cyclic Brinell indentation scratching technique [15]. A 2 x 2 mm$^2$ area with a dislocation density of ~$10^{13}$ m$^{-2}$ in the near surface (penetrating with a depth of ~100 μm [9]) was created in the dislocation-engineered samples by scratching a 2.5 mm hardened steel Brinell sphere under a load of 10 N over the sample surface making 10 passes over every point of the plastic zone. This yields a dislocation density that is three orders of magnitude higher compared to the as-received samples (~$10^{10}$ m$^{-2}$) [15]. The surface features for the as-received and dislocation-engineered samples were captured in an optical microscope (Zeiss Axio Imager2, Carl Zeiss AG, Oberkochen, Germany) with the circularly polarized light–differential interference contrast mode (C-DIC). They are depicted in **Fig. S1(A)** and **Fig. S2(A)** in the Supplementary Materials. Additionally, the dislocation-engineered area was scanned with 3D laser confocal scanning microscopy using a LEXT OLS 4000 (Olympus IMS, Waltham, USA), equipped with a DIC prism to give contrast to subtle height changes on the surfaces of the samples. A line scan, visualizing the surface condition after dislocation engineering, is depicted in **Fig. S3** in the Supplementary Materials.

For testing the dielectric breakdown strength, samples were placed in the setup in **Fig. 1(A)**, which was purposely constructed for these experiments. This setup consists of two spring-loaded pins connected to a TREK Model 20/20C high-voltage amplifier with a maximum voltage of 20 kV. To avoid dielectric breakdown through the air, a ceramic oil vessel was placed around the lower pin and filled with silicone oil. The samples were placed in the oil, balanced on the lower pin, and contacted with the upper pin. Then, the voltage was increased with 0.2 kV/s until the sample shattered. Voltage and current were recorded, and the samples were imaged using optical microscopy after the dielectric breakdown event for comparison, as presented in **Fig. S1(B)** and **Fig. S2(B)** in the Supplementary Materials.

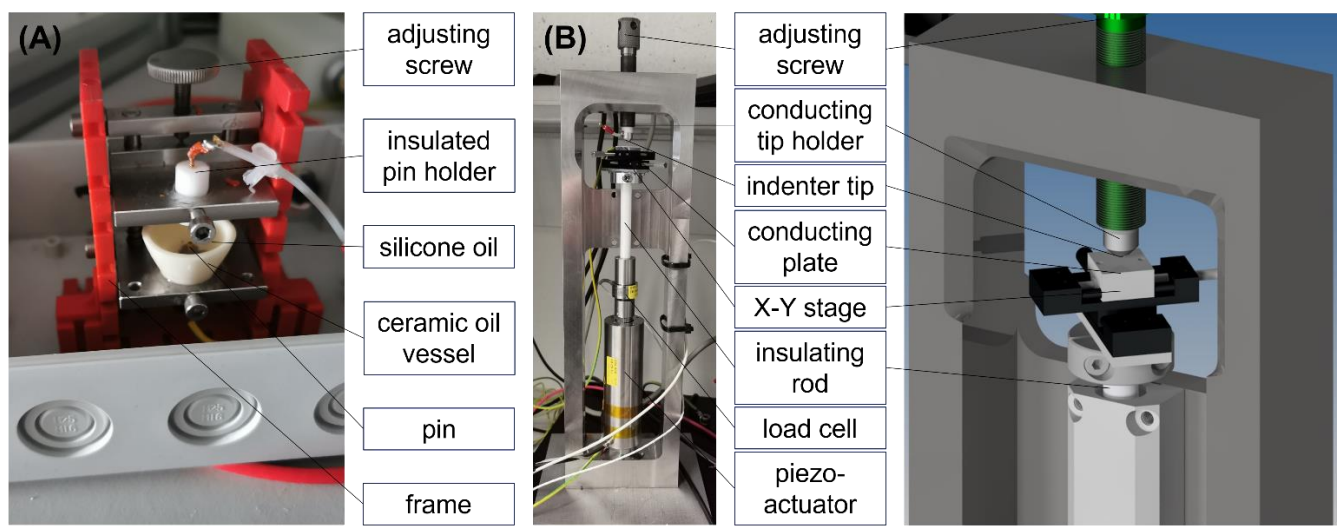

*Figure 1. (A) Image of the setup used for dielectric breakdown experiments. The sample is held between two pins attached to a high voltage source while being submerged in silicone oil held in a ceramic vessel. (B) Image and schematic of the setup designed for the electric field-assisted spherical indentation experiments. The sample is placed under the conducting indenter tip and on the conducting plate, which are both connected to a high voltage source.*

**2.3 Brinell indentation with electric field**

To investigate the effect of an electric field on the plastic deformation behavior of SrTiO$_3$ during Brinell indentation, an indentation setup coupled with electric field was designed, as depicted in **Fig. 1(B)**. The samples were placed on a metallic base plate connected to a movable stage enabling the selection of the indentation site. The indenter consists of a hardened steel sphere with a diameter of 2.5 mm, attached to a conducting holder for the indenter. Both the metallic base plate and the indenter holder are connected to the same high-voltage amplifier used in the dielectric breakdown experiments. The indenter is attached to an adjusting screw for manually applying a pre-load on the sample, to ensure proper operating conditions for the load cell (Precision Miniature Tension and Compression Load Cell Model 8431-5200, burster GmbH & Co. KG, Gernsbach, Germany) and the PID-control of the piezo actuator (P-216 PICA Power Piezo Actuators, Physik Instrumente, Karlsruhe, Germany) through which the load is applied. The load measured by the load cell is acquired by a data acquisition module (HBM MGCplus TG009E with an ML55B amplifier module, Hottinger, Brüel & Kjær, Darmstadt, Germany). The piezo actuator is controlled by an amplifier (E-481 PICA Piezo High-Power Amplifier/Controller, Physik Instrumente, Karlsruhe, Germany). The force from the piezo actuator is transferred through an insulating rod to shield the load cell and actuator from the high voltage applied to the sample.

The experimental process was as follows: the sample was placed on the metallic base plate, an indentation site was chosen by moving the stage, then the adjusting screw was carefully lowered until a pre-load of 5 N was reached. Subsequently, a load of 15 N was approached with a rate of 1 N/s by the extension of the piezo-actuator and held for 10 s. Afterwards, the load was lowered to 5 N again with 1 N/s unloading rate. This series of operations is termed as 1 load cycle. This was tested under three conditions of external field: with the voltage source turned off (0 kV), with +2 kV applied, and with -2 kV applied between the indenter tip and the metallic base plate. In addition to single load cycles, the experiments were also repeated with 10 load cycles of loading under all three different field conditions, as well as with the holding period under load extended to 1 h. Load, applied voltage, and the current through the samples were continuously monitored during the tests.

After the experiments, the samples were chemically etched to reveal the dislocation etch pits, with the etching procedure described elsewhere [14]. The etched samples were imaged using the same C-DIC mode in the Zeiss optical microscope, as well as 3D laser confocal scanning microscopy, using the same setups described above.

**3. Results and analyses**

**3.1 Dielectric breakdown**

**Figure 2(A)** shows an exemplary plot of the electric field across a sample, as well as the current measured during the experiment. For each test, the electric field increased linearly until the dielectric breakdown strength of the tested specimen was reached, at which point the voltage across the sample, and therefore the applied electric field, decreased rapidly. No detectable current was captured during any test up until reaching the dielectric strength, which was signified by a sudden increase in current flow. The maximum electric fields measured in these experiments were then used as the dielectric breakdown strengths of the tested specimens.

The dielectric breakdown strengths of the as-received as well as the dislocation-engineered $SrTiO_3$ samples were compared for their Weibull failure probability. The corresponding Weibull plots are presented in **Fig. 2(B)**, where the blue curve represents the failure probabilities of the as-received samples and the red curve those of the dislocation-engineered samples. For the as-received samples, the Weibull plot is defined by a specific breakdown strength $E_0$=30.3 kV/mm and a Weibull modulus m=12.2. The measured values were between 24.6 and 33.6 kV/mm, which agrees with reported values of the breakdown strength of single crystalline $SrTiO_3$ [16]. Scratching the samples for 10 passes with a spherical indenter tip however reduces both specific breakdown strength and Weibull modulus to $E_0$=24.3 kV/mm and m=4.4. While the maximum value of the dielectric breakdown strength remained rather unchanged with 31.9 kV/mm, one of the dislocation-engineered samples failed already at 13.5 kV/mm. The scratching procedure therefore appears to have lowered the dielectric breakdown strength and reliability of $SrTiO_3$ single crystals.

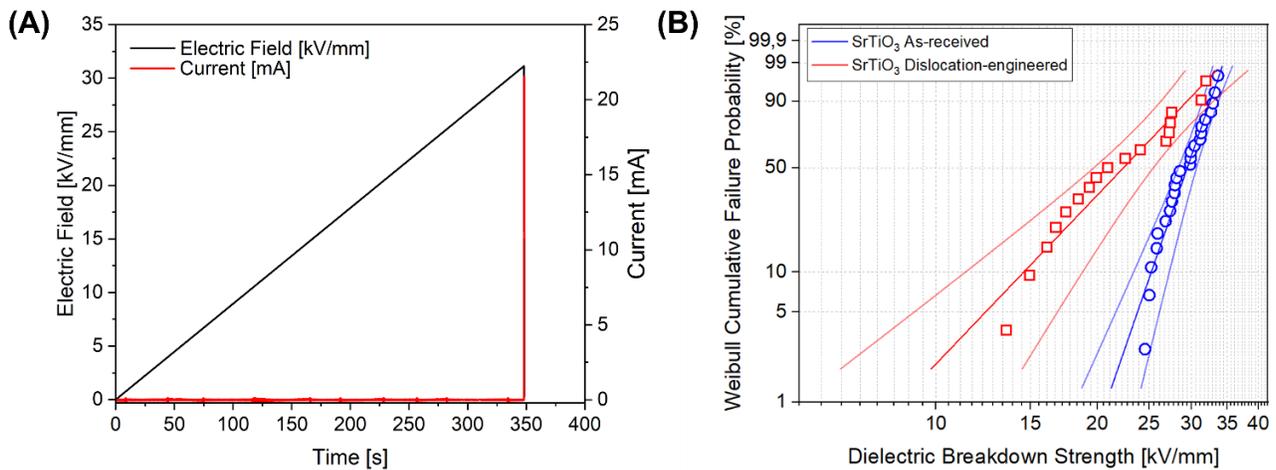

***Figure 2.*** *(A) Electric field and current during the dielectric breakdown. The field increases linearly until the sample's breakdown strength is reached, after which it decreased rapidly. Simultaneously, the current increases sharply. (B) Weibull cumulative failure probability for as-received and dislocation-engineered $SrTiO_3$. After deformation, the dielectric breakdown is reached at lower electric field strengths.*

## 3.2 Brinell indentation under electric field

Brinell spherical indentation experiments with an applied electric field were conducted with 1 and 10 cycles of mechanical loading and three different electric field conditions: 0 kV/mm as a reference, and +2 and -2 kV/mm for investigation of the electric field effect on the deformation with opposite polarities. Here, the field strength of 2 kV/mm was chosen to safely avoid dielectric breakdown of the sample during deformation (as presented in the last section) and of the air surrounding the sample. Furthermore, this field strength was previously successfully utilized to assist the deformation in NaCl and ZnS single crystals [6, 7]. The applied load and voltage for a 10-cycle indentation experiment with an electric field of +2 kV/mm are presented in **Fig. 3(A)**, for a 1-cycle experiment the corresponding graphs are displayed in the supplementary **Fig. S4**.

Additionally, **Fig. 3(B)** depicts the measured current through the sample, where no current above the background was detected. This was expected as nominally undoped $SrTiO_3$ used in these tests is considered insulating with a band gap of 3.2 eV [17]. While an applied electric field can increase the conductivity of $SrTiO_3$, it would require 200 kV/mm to increase the conductivity by a factor of 2 [18], which is 100 times the applied electric field in this experiment. Therefore, the increase to the samples' conductivity was considered negligible, as the mobility of charge carriers in $SrTiO_3$ is limited at room temperature [18-20]. For this reason, influences of Joule heating were not considered here.

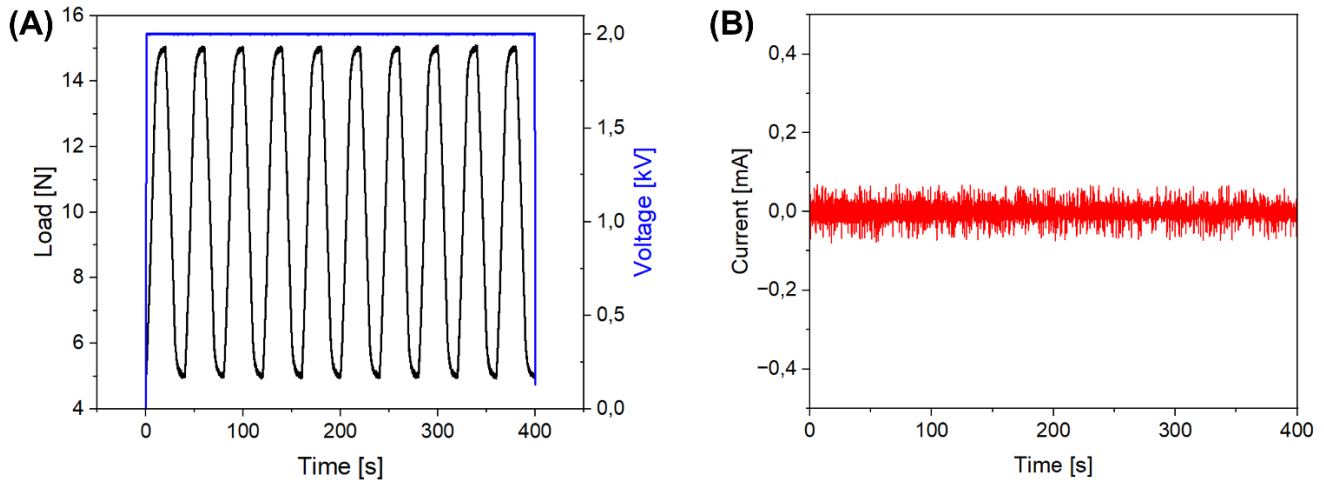

*Figure 3. (A) Applied load and voltage during 10 cycles of electrostatic indentation. (B) Current through the sample measured during 10 cycles of electrostatic indentation.*

After the deformation under the electric field, the sample was chemically etched to reveal the generated dislocations on the sample surface. The optical microscopy images of the indents are presented in **Fig. 4(A)**. The etch pit rosettes around the indent imprints without the applied electric field are in excellent agreement with previously reported surface slip and dislocation etch pits arrangements around Brinell indents, both in size as well as in length and number of slip traces [21]. When compared to the indents with the applied electric field, no detectable changes were observed regarding the size of the indents, the number and length of slip traces, and the distribution of dislocation etch pits. This is the case for both 1 and 10 cycles of indentation, as well as for applying +2 or -2 kV/mm. For comparison of the imprints, the plastic zone size was evaluated by fitting a circle to envelope the outermost visible etch pits [22]. For 1 cycle of indentation the average plastic zone had a radius of 154±5 µm, 154±4 µm, and 152±7 µm for the tests with 0, +2, and -2 kV/mm, respectively. Similar results were obtained for 10 cycles of indentation, where the plastic zone radius averaged at 184±3 µm, 183±8 µm, and 181±5 µm for the tests with 0, +2, and -2 kV/mm, respectively.

For further comparison, line scans across the indent imprints were taken using laser confocal microscopy, as plotted in **Fig. 4(B)**. One cycle of spherical indentation (red) produces indents with a depth of ~200 nm, after 10 cycles (blue) this depth increases to ~300 nm. It is evident that applying an electric field of 2 kV/mm does not alter the size, nor depth of the indents significantly, independent of the field's polarity.

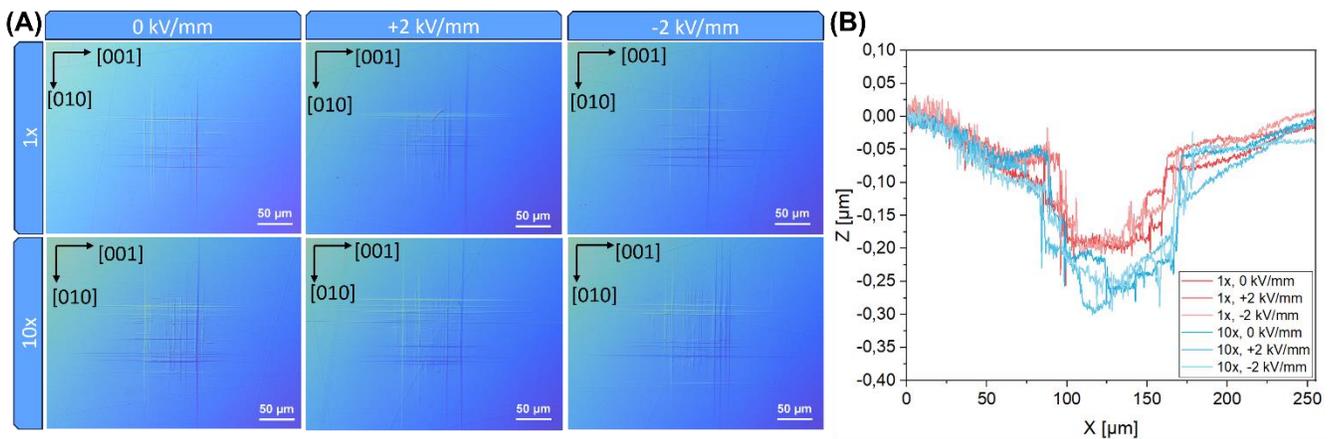

*Figure 4. (A) Differential interference contrast microscopy images of the etched sample surface after 1 and 10 cycles of spherical indentation with 0, +2, and -2 kV/mm of applied electric field. (B) Line scans from laser confocal microscopy of the indents.*

### 3.3 Load-holding electrostatic Brinell indentation

In the previous experiments, the time for which the electric field was applied was short to intermediate (30 to 400 s). To investigate the effect of longer exposure times of stress and electric fields, load-holding experiments were conducted, in which the sample was loaded same as for the 1-cycle indent, but the applied load and electric field were held for 1 h (3600 s). The plots of the applied voltage and current as well as the measured current are presented in the supplementary material **Fig. S5**. Again, no current flow was detected during the period of applied electric field.

After chemical etching, the etch pit rosettes around the imprints were revealed. Optical microscopy images and laser confocal microscopy line scans are presented in **Fig. 5**. Again, the size of the plastic zone was evaluated by enveloping the outermost etch pits in a fitted circle. The radius of the plastic zone had average values of 167±14 µm, 160±8 µm, and 162±16 µm for the tests with 0, +2, and -2 kV/mm, respectively. Holding the load for 1 h produced a larger imprint than a 10 s holding time, which clearly evidences the additional movement of dislocations under this condition (analogous to indentation creep under constant load although the stress is strictly speaking not constant). The size as well as the depth of the indent is more comparable to 10 cycles of indentation presented in the previous section. Application of the electric field, regardless of its polarity, exhibited no clear influence on the size or depth of the indent imprints. Furthermore, the overall arrangement and distribution of etch pits remained the same with an applied voltage to the sample. Therefore, an external electric field with the current field strength does not appear to have an influence on the motion of dislocations in single-crystalline $SrTiO_3$ at meso-scale.

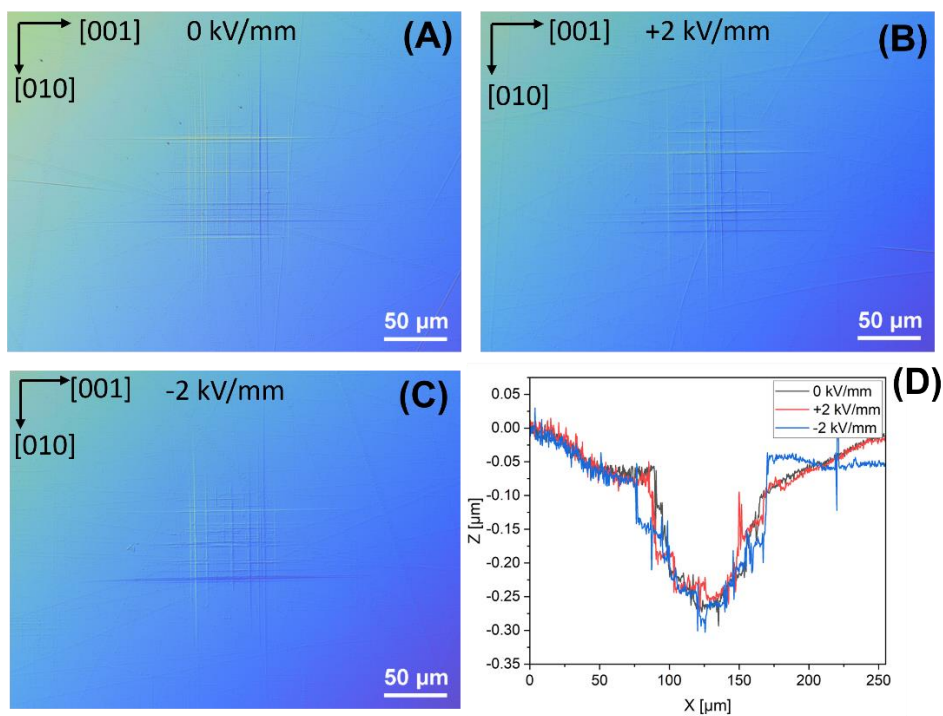

*Figure 5. Differential interference contrast microscopy images of the etched sample surface after 1 hour of load holding with (A) 0, (B) +2, and (C) -2 kV/mm of applied electric field and (D) line scans from laser confocal microscopy of the indents*

## 4. Discussion

### 4.1 Influence of dislocations on dielectric breakdown

Increasing the dislocation density of SrTiO$_3$ by 10 passes of scratching with a spherical indenter tip effectively lowered its dielectric breakdown strength, accompanied by a lower Weibull modulus. As various factors can influence the dielectric breakdown strength of a material [23, 24], we attempted to keep the testing parameters comparable by using samples from the same batch and conducting the dielectric breakdown experiments on the same day, to exclude changes in sample composition, sample history, sample thickness, temperature, or humidity. Furthermore, pin contacts were chosen to minimize the probability of encountering major flaws, like microcracks or voids, in the bulk of the samples. The remaining key aspects relevant for the observed decrease in dielectric strength are discussed in the following.

By scratching a polished SrTiO$_3$ surface for 10 passes with a spherical indenter tip the near-surface region (~100 µm in depth) dislocation density is increased to ~$10^{13}$ m$^{-2}$ [15]. The produced scratch tracks have a depth comparable to the depth of the Brinell indents, as presented by the line profiles in **Fig. 4(B)**. This treatment can mechanically introduce several extrinsic factors that have been reported to affect the dielectric breakdown strength. Besides dislocations, it is conceivable that sliding a hardened steel sphere across a brittle ceramic crystal surface could introduce near-surface damage. It has been reported that the presence of voids, microcracks and other flaws in a dielectric material can effectively lower its dielectric breakdown strength through local field enhancement [25, 26]. Furthermore, the surface roughness is increased when scratching a larger area, as was the case in these experiments, where the scratching procedure introduced a periodic surface curvature (see **Fig. S2**). An increase in surface roughness can lead to concentration of the electric field, which in turn can lower the externally observed breakdown strength of a material [27, 28]. Lastly, it has been observed that scratching introduces residual compressive stresses into the sample, which can inhibit crack initiation and propagation [29]. As fracture is a part of the dielectric breakdown process, hindering crack initiation and propagation could shift the dielectric breakdown strength to higher values [30].

However, it remains unclear to what degree these factors have influenced the breakdown measurements in this study. While pile-ups of dislocations have been reported to be crack precursors in the form of Zener-Stroh pre-cracks [31], no visible cracks of any size were detected using optical microscopy, as presented in **Fig. S2**. The transparency of the tested SrTiO$_3$ samples allowed for the shifting of the focus plane into the sample surface during optical microscopy, where no cracks were observable as well. Furthermore, the introduction of dislocations into SrTiO$_3$ has been reported to increase the material's indentation fracture resistance [32], which opposes the presence of microcracks. It is therefore unlikely that the process of generating dislocations has introduced other flaws responsible for lowering the breakdown stress. As for the increased surface roughness, it can be seen from the laser line scan in **Fig. S3**, that the scratching procedure introduced "hills and valleys" with a depth of ~0.4 µm and a periodicity of 100 µm. Even though a pin contact was used, the pin would still contact the sample over several of these hills, where a concentration of the electric field could occur through local field enhancement. This could be one of the reasons for the lowered dielectric breakdown strength. Moreover, as the residual stresses (due to scratching [29]) contributed to the increase in indentation fracture resistance, it is therefore unlikely the residual compressive stress caused the decrease of the breakdown strength. To avoid both of those influencing factors, future experiments of this kind should consider an additional polishing and annealing step after the scratching procedure, in order to minimize surface roughness and compressive residual stresses, respectively.

Apart from the purely mechanical factors, dislocations have been shown to influence both electrical and thermal conductivity of $SrTiO_3$. As both factors can have implications on the dielectric breakdown behavior, they will be discussed in the following two parts.

First, from an electromechanical point of view, dislocations in $SrTiO_3$ have been discussed for their potential influence on the electrical conductivity [33]. An increase in electrical conductivity may decrease the dielectric breakdown strength and increase the degree of uncertainty in the breakdown probability, due to local Joule heating or the accumulation of charge carriers, which, when accelerated, can be the starting point of the dielectric breakdown process [34]. Dislocations in $SrTiO_3$ have been argued to be able to increase its electrical conductivity, either through an increase in the density of mobile charge carriers or by allowing for faster diffusion along the dislocation lines [33]. It is, however, unclear if this can explain the lower breakdown strength observed here. As elaborated earlier, no current flow was detected, even when applying 2 kV for one hour, as depicted in **Fig. S5**. Furthermore, Porz et al. [33] demonstrated that the influence of dislocations (mechanically induced, without additional reduction treatment of the samples) on the electrical conductivity of $SrTiO_3$ is small, if not non-existent. Additionally, the dislocations introduced by sphere scratching are not uniformly distributed throughout the sample but are confined within the first few hundreds of micrometers from the surface [15, 35], therefore creating no conductive path through the sample. If there is an influence on the dielectric breakdown behavior, it would only become relevant during the breakdown event, when the temperature suddenly rises locally.

This leads to discussion on the second part that concerns the thermomechanical aspect. It has been recently reported that a high dislocation density ~$10^{15}$ m$^{-2}$ can significantly reduce the thermal conductivity by up to ~50% in nominally undoped single-crystal $SrTiO_3$ [36]. This is due to the scattering of phonons when the distance between dislocations reaches the phonons' mean free path, as in the case with high-density dislocations [36]. Furthermore, it was experimentally demonstrated that conduction along dislocations can increase the local temperature more than conduction through an undeformed volume [37]. The dielectric breakdown process can, however, be a thermal runaway process, in which Joule heating increases the electrical conductivity and the increased conductivity in turn generates more Joule heating [24, 38]. If the heating of the sample is aided and heat dissipation is hindered by the presence of dislocations, it is likely that dislocations would effectively lower the dielectric breakdown strength of $SrTiO_3$.

As there are many factors that may influence the breakdown behavior in the conducted experiments, it is as of now not possible to discern the contribution of each influencing factor on the lowered dielectric breakdown strength. While the increased presence of microcracks or other flaws after the scratching procedure can be excluded from the discussion, the changes in surface roughness as well as the residual compressive stresses should be considered as influencing factors from the mechanical point of view. As the extent of the increase in electrical conductivity due to dislocations in $SrTiO_3$ at room temperature is marginal, its influence on the dielectric breakdown behavior remains debatable. The changed thermal behavior of $SrTiO_3$ upon increasing the dislocation density could however be influencing the breakdown strength. Future experimental design shall consider additional processing procedures of the samples to deconvolute these influencing factors.

**4.2 Electric field effect on the deformation of $SrTiO_3$**

Here, we discuss the results of the spherical indentation with an applied electric field. For this, a closer examination of the dislocation core and its arrangement in the lattice is necessary. The room-temperature mobile dislocations in $SrTiO_3$ belong to the {110}<110> systems. However, during spherical indentation on the (001) surface only the {110} planes that are 45° inclined to the indented surface are activated due to the stress state under the spherical tip [21, 39]. Note, that these slip planes intersect with the surface and produce the slip traces that are arranged vertically and horizontally as in **Figs. 4, 5**. These slip planes also give the maximum Schmid factor of 0.5 if under uniaxial compression in the <001> direction. Therefore, the dislocation half-loop, that is introduced under the indenter, has its screw components going into the surface, and its edge components at the bottom of the loop [39], as depicted in **Fig. 6**. To be aided in its motion by an electric field, the dislocation needs to be electrically charged, and the electric field needs to exert a force onto the charged dislocation that points in the direction of glide [6]. As the screw components of dislocations in $SrTiO_3$ are considered charge-neutral [10], the edge dislocation core will be discussed more closely in the following.

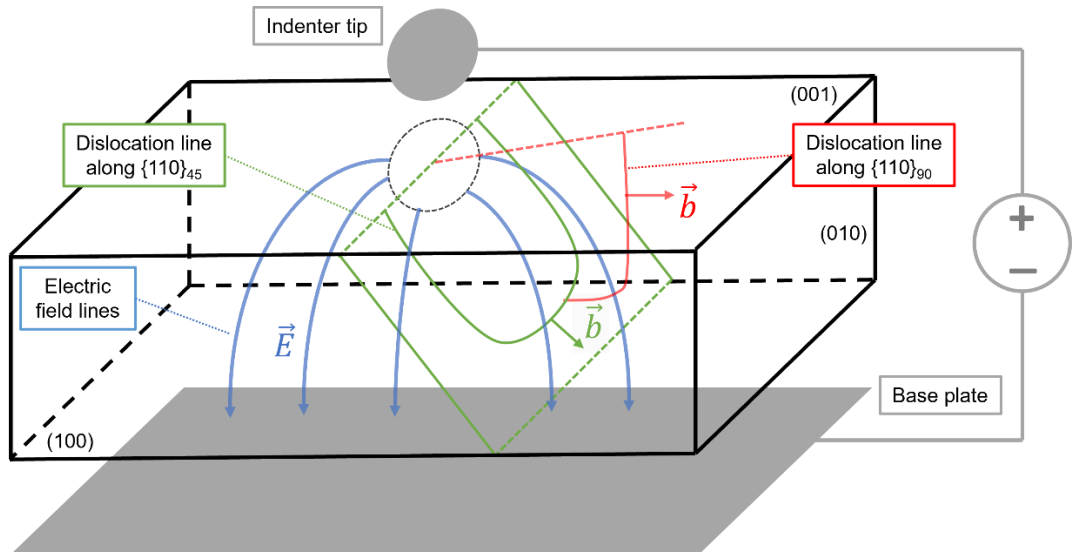

*Figure 6. Schematic of dislocation half-loops in a SrTiO$_3$ single crystal introduced by spherical indentation. In green, a dislocation on the {110}$_{45}$ plane (angled at 45° to the indented surface) is depicted, red shows a dislocation on the {110}$_{90}$ plane (angled at 90° to the indented surface). The indenter tip is connected to the base plate by a voltage source; the blue arrows represent the electric field lines as an indicator for the applied electric field.*

To estimate the effect of an electric field on the glide of an edge dislocation in SrTiO$_3$, the charge states of the dislocation cores can play a critical role. The charge state is strongly dependent on the exact atomic arrangement of the dislocation core in SrTiO$_3$, which has been an object of debate [10, 11]. It was reported by Klomp et al. [10] that the edge dislocation is either charge-neutral, undissociated and therefore immobile; or mobile, dissociated and charged either positively or negatively by an amount of two elementary charges per line vector of dislocation. More recently, however, Hirel et al. [11] presented a dissociated charge-neutral dislocation core, which is energetically more favorable than the previously reported core structures. Furthermore, it was stated that the charge-neutral core could acquire positive charges by interacting with oxygen vacancies that are readily present in nominally undoped SrTiO$_3$. In the simulation experiment the positively charged dislocation stayed mobile, although the shear stress required for dislocation motion increased from 70 MPa to 240 MPa [11].

On the one hand, as discussed above, the dislocations introduced by the Brinell spherical indentation, as depicted in **Fig. 6**, appear mostly screw types and hence charge-neutral. This would explain the results of the electric field-assisted indentation experiments. Charge-neutral dislocations cannot be affected by the applied electric field, therefore no changes in the deformation behavior have been observed. On the other hand, if we consider the fraction of edge dislocations that are positively charged, one can estimate the additional force on the dislocation exerted by an electric field. For this, we consider the forces acting on a moving dislocation segment. The mechanically applied force per unit length of dislocation $f_m$ can be expressed as [6]:

$$f_m = \tau \cdot b \tag{1}$$

where $\tau$ is the applied shear stress and $b$ is the Burgers vector. With the necessary shear stress of 240 MPa [11] for the positively charged dislocation and the Burgers vector of $a$/2 [110], where $a$ is the lattice constant of 3.905 Å [40], the force per unit length required for motion becomes 0.066 N/m. The electrostatically applied force per unit length of dislocation $f_e$ can be expressed as [6]:

$$f_e = \phi \cdot E \cdot q_l \tag{2}$$

Here, $E$ is the applied electric field, $q_l$ is the charge per unit length of dislocation and $\phi$ is a geometrical orientation factor, as the force exerted by the electric field needs to point into the glide direction. In the case of the experiments conducted, the electric field is applied from electrodes at the top and bottom of the sample, the glide direction, however, is at 45° to the sample surfaces, as demonstrated earlier. As a simple approximation to the complex electric field condition of a spherical electrode on one side, a base plate on the other side, and the sample in between, the geometrical factor was chosen as 0.5. The applied electric field was 2 kV/mm and the charge per line vector was reported as the charge of one oxygen vacancy, two positive elementary charges with a unit line vector of $a$[010]. This results in a force per unit length of 0.821 mN/m, or 1.2 % of the force required to move a dislocation. In other words, at an electric field of 2 kV/mm the additional force on a positively charged dislocation line is negligibly small. Conversely, to move a charged edge dislocation through $SrTiO_3$ by an electric field alone, a field strength of 243 kV/mm would be necessary, which is ~7 times larger than the maximum breakdown strength measured in this work for the meso/macroscale samples. This simple mathematical description demonstrates that to efficiently assist the charged dislocations in their motion, a significantly larger electric field would be required, which in turn increases the likelihood of dielectric breakdown, especially in the presence of pre-engineered dislocations.

This raises the question why comparable electric fields were able to assist the plastic deformation in NaCl and ZnS single crystals but not in $SrTiO_3$? In NaCl it was demonstrated how dislocations can sweep up cation vacancies to acquire a charge [6], with a maximum achievable number of one vacancy every second available cation position along the dislocation line [41]. The maximum charge per unit length in this case would be half of an elementary charge with a line vector of $a$[010], which is a quarter of the charge estimated for the positively charged dislocations in $SrTiO_3$. However, the critical resolved shear stress required to move a dislocation through NaCl at room temperature is much lower at around 3 MPa [42]. With a lattice parameter of 5.59 Å and a Burgers vector of $a$/2 [110], the force per unit length required to move a dislocation becomes 1.185 mN/m according to Equation (1). The force applied to the charged dislocation through the electric field becomes 0.143 mN/m with Equation (2), which is roughly 12 % of the stress required for motion, large enough to assist in the deformation of NaCl.

The same procedure can be conducted for dislocations in ZnS. With the critical resolved shear stress (from compression experiments in darkness [43]) and experimentally determined charges per unit length of partial dislocations [7], the stress per unit length required to move a dislocation becomes 2.697 mN/m and the force applied to the charged dislocation through the electric field becomes 0.176 mN/m, or roughly 6.3 % of the force required for motion of the dislocations. This is peculiar, as it was recently shown by Li et al. that dislocations in ZnS can be moved by an applied electric field alone [8]. However, this discrepancy may be explained by consideration of the investigated length scale. In the study by Li et al. the samples were cut from the bulk by the focused ion beam method resulting in very thin lamella, in which the electric field was concentrated, achieving field strength 12 times larger compared to this study. Furthermore, at these length scales, image forces may effectively affect the dislocations and assist in their motion. This showcases the influence of the length scale on the electroplastic effect in ceramics, namely, the electric field manifests itself more significantly at smaller scales.

For $SrTiO_3$ similar considerations may be worthwhile, as thinner samples have larger dielectric breakdown strengths [34]. With thin films of thickness around $10^{-4}$ mm (or 100 nm) the dielectric breakdown strength of $SrTiO_3$ increased to roughly 300 kV/mm, which would be enough to move positively charged dislocations without an external field. At these length scales, the mobility of oxygen vacancies can also play a role. For nominally undoped $SrTiO_3$, a general description of the temperature dependent self-diffusivity $D_V$ can be expressed as [18, 19]:

$$\ln[D_V/cm^2 s^{-1}] = -4.96 - \frac{0.67\ eV}{k_B T} \qquad (3)$$

Here, $k_B$ is the Boltzmann constant, and $T$ is the temperature. With a temperature of 300 K this yields a diffusivity of $4\times10^{-14}$ cm$^2$/s. With this the average distance *l* travelled by the vacancies can be estimated through:

$$l = \sqrt{D_V t} \qquad (4)$$

Assuming a time *t* of 3600 s as in the load holding experiments, this results in a distance of $1.2\times10^{-4}$ mm, which means that vacancies in such thin films would be able to diffuse over sufficient distance towards dislocations. While the presence and motion of oxygen vacancies can have a significant influence on the behavior of dislocations in $SrTiO_3$ [22, 44], this highlights again the importance of the length scale when investigating the electroplastic effect in ceramics.

The results of this study have two implications for dislocation-engineered electroceramics. Firstly, the maximum voltage, which can be applied to electroceramics with large dislocation densities, may be lower due to the previously discussed factors that lower the dielectric breakdown strength. While some extrinsic factors like the surface roughness, residual stresses, or the presence of flaws from the introduction of dislocation may be mitigated by proper post-processing of the samples, other intrinsic factors like the increased electrical conductivity or the reduced thermal conductivity, which may even be desirable for the functional properties, cannot directly be avoided. These aspects need to be accounted for when designing dislocation-engineered devices. Secondly, it can be argued that once dislocations are engineered into a perovskite ceramic, they exhibit a strong stability against externally applied voltages in the sense that the bulk material would fail long before the dislocations are moved by the electric field. While at very small length scales, dislocations may be influenced by a larger electric field and the motion of oxygen vacancies, for the meso- and bulk scale investigated here the dislocations remain stable against applied electric fields.

## 5. Conclusion

Room-temperature dislocation plasticity in nominally undoped single crystal $SrTiO_3$ under an external electric field was investigated, with the goal of testing the feasibility of electric field-assisted plastic deformation and the stability of dislocation under external electric fields at mesoscale. Dielectric breakdown strength measurements demonstrate a lower dielectric breakdown strength in the dislocation-rich samples (~24 kV/mm) compared to the reference samples (~30 kV/mm). Besides extrinsic factors such as an increased surface roughness or residual stresses from the scratching procedure, intrinsic factors including changes in electrical conductivity and thermal conductivity from the increase in dislocation density are considered to explain the lower resistance to external electric fields in dislocation-rich samples. Brinell indentation tests with an applied electric field lower than the dielectric breakdown strength was further conducted to investigate its impact on the dislocation generation and the feasibility of utilizing an electric field in the deformation of perovskite ceramics. At mesoscale, no observable changes in the plastic zone (for both short and 1-hour load hold or cyclic indents) were observed despite an applied field strength of 2 kV/mm. This behavior is explained through the charge state of the dislocations in $SrTiO_3$. While screw dislocations are neural and edge dislocations may carry a positive charge, the force exerted by the electric field of this magnitude is negligible. Our findings suggest that electric fields with low/intermediate strength may not enhance the introduction of dislocations into perovskite $SrTiO_3$ at mesoscale. Nevertheless, the stability of the pre-engineered dislocations under electric field is clearly demonstrated.

**Acknowledgement**

A. F. and X. F. acknowledge the funding from the European Research Council (ERC Starting Grant, Project MECERDIS, grant No. 101076167). Views and opinions expressed are, however, those of the authors only and do not necessarily reflect those of the European Union or the European Research Council (ERC). Neither the European Union nor the granting authority can be held responsible for them. O.P. acknowledges the financial support by DFG (No. 414179371).

# Supplemental Materials for

**Dislocation response to electric fields in strontium titanate: A mesoscale indentation study**

Alexander Frisch[1]*, Daniel Isaia[2], Oliver Preuß[2], Xufei Fang[1]*

[1]Institute for Applied Materials, Karlsruhe Institute of Technology, Karlsruhe, Germany
[2]Department of Materials and Earth Sciences, Technical University of Darmstadt, Darmstadt, Germany

*Corresponding authors: alexander.frisch@kit.edu (AF); xufei.fang@kit.edu (XF)

**This file contains the following information:**

**Figure S1.** Differential interference contrast microscopy images of an as-received $SrTiO_3$ single crystal sample **(A)** before and **(B)** after dielectric breakdown. The sample broke into four pieces along <100> and <010> directions.

**Figure S2.** Differential interference contrast microscopy images of a dislocation-engineered $SrTiO_3$ single crystal sample **(A-C)** before and **(D-F)** after dielectric breakdown. **(B, C)** Close-up images of the introduced dislocation structures. **(D-F)** Sample after dielectric breakdown and subsequent chemical etching. Highly crystallographic crack patterns are visible.

**Fig. S3.** Laser line scan across the dislocation-engineered area in **Fig. S2**. The numerical values of the axes represent the measured distances in x- and z-directions.

**Fig. S4.** (A) Applied load and voltage during 1 cycle of electrostatic indentation. (B) Current through the sample measured during 1 cycle of electrostatic indentation.

**Fig. S5.** (A) Applied load and voltage during electrostatic indentation with 1 hour of load holding. (B) Current through the sample measured during electrostatic indentation with 1 hour of load holding.

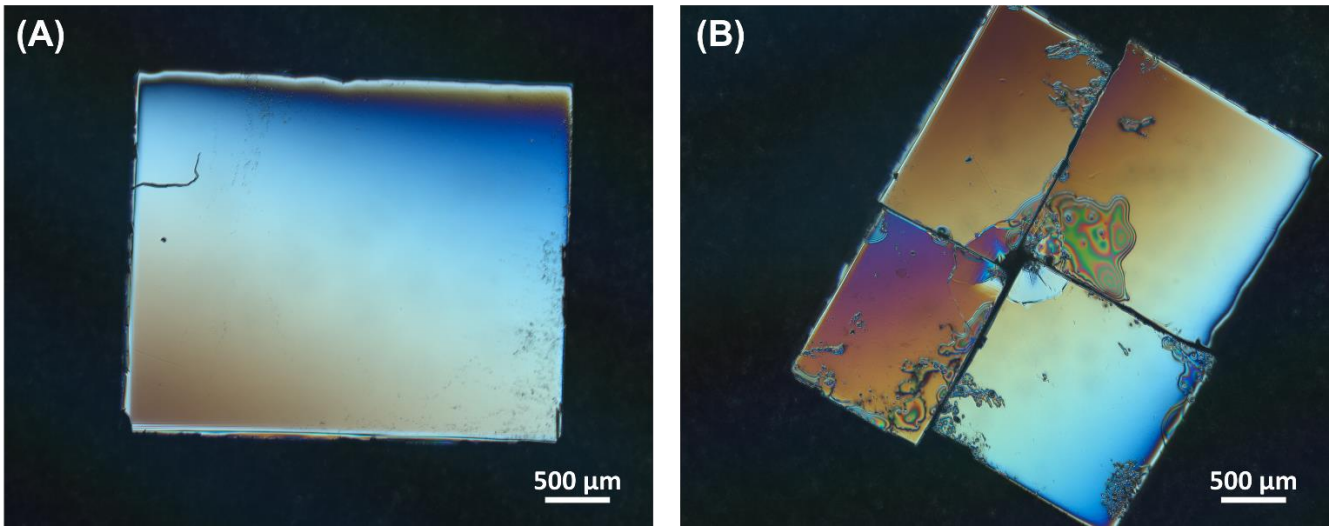

**Figure S1.** Differential interference contrast microscopy images of an as-received $SrTiO_3$ single crystal sample **(A)** before and **(B)** after dielectric breakdown. The sample broke into four pieces along <100> and <010> directions.

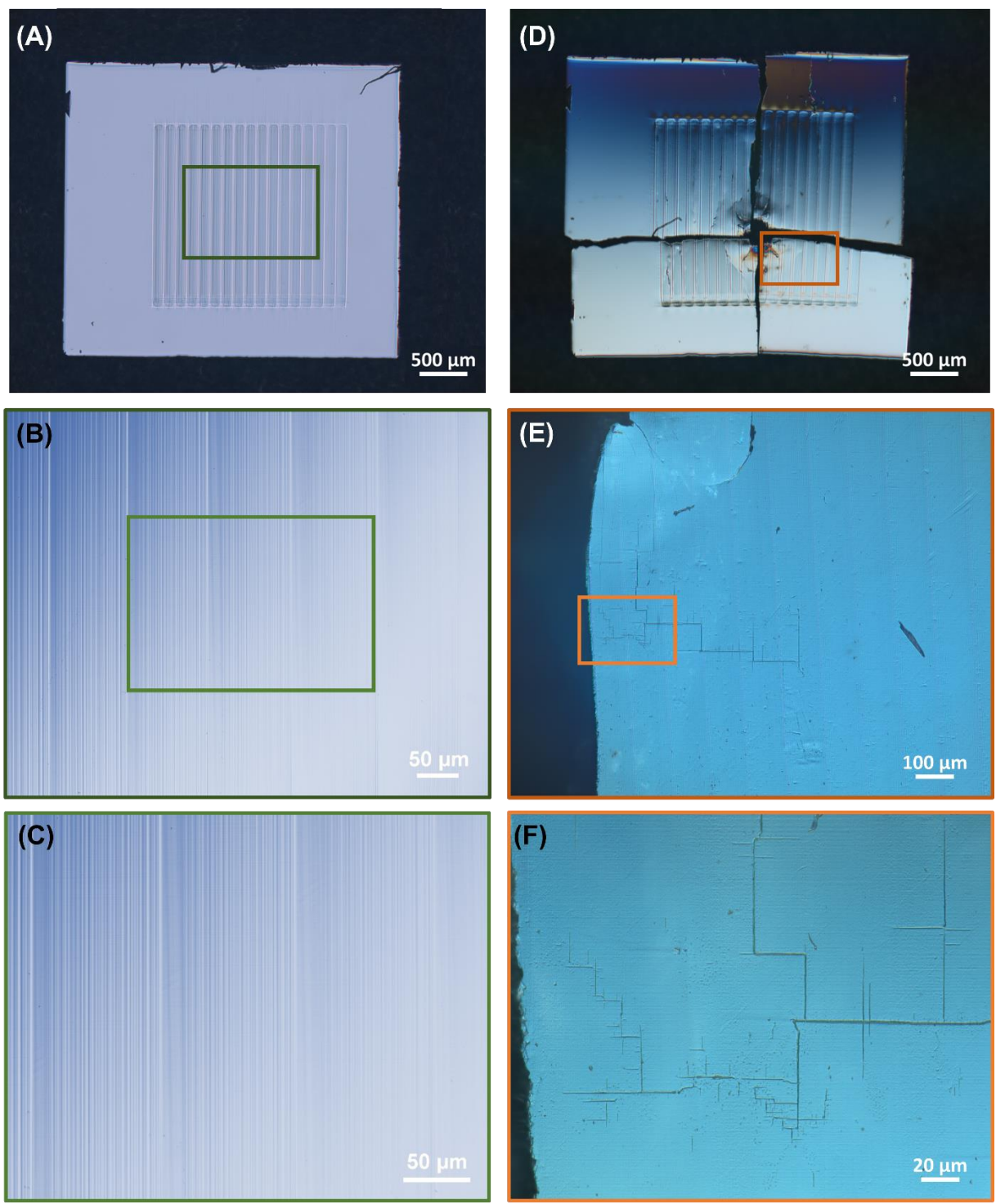

**Figure S2.** Differential interference contrast microscopy images of a dislocation-engineered SrTiO$_3$ single crystal sample **(A-C)** before and **(D-F)** after dielectric breakdown. **(B, C)** Close-up images of the introduced dislocation structures. **(D-F)** Sample after dielectric breakdown and subsequent chemical etching. Highly crystallographic crack patterns are visible.

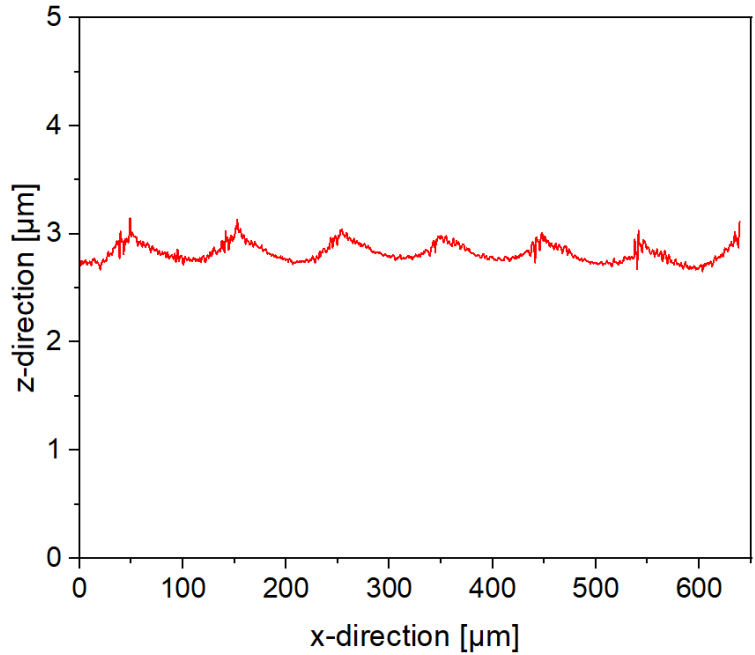

**Figure S3.** Laser line scan across the dislocation-engineered area in **Fig. S2**. The numerical values of the axes represent the measured distances in x- and z-directions.

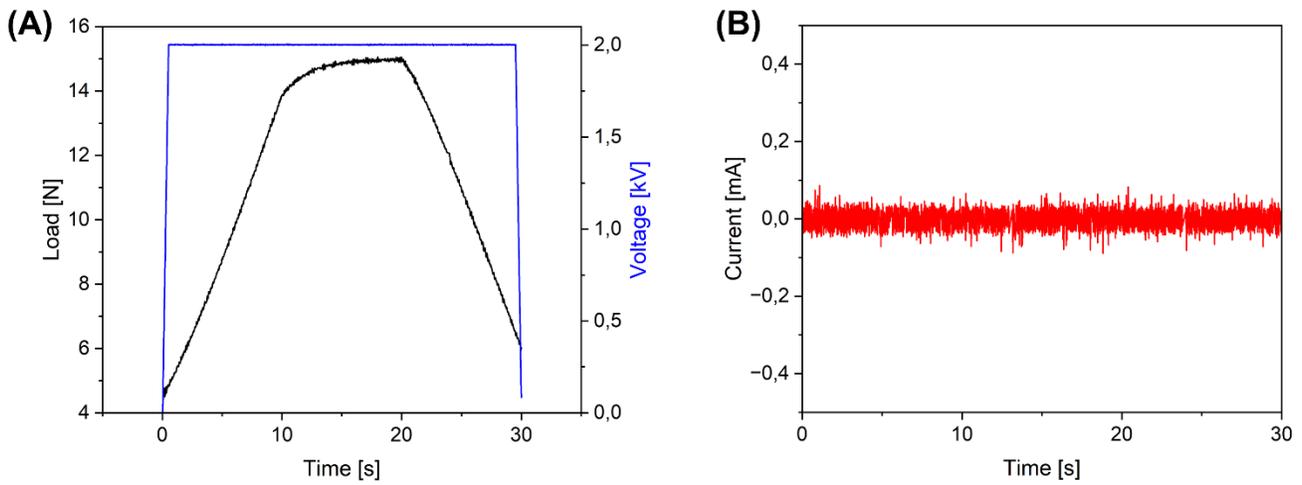

**Figure S4.** (A) Applied load and voltage during 1 cycle of electrostatic indentation. (B) Current through the sample measured during 1 cycle of electrostatic indentation.

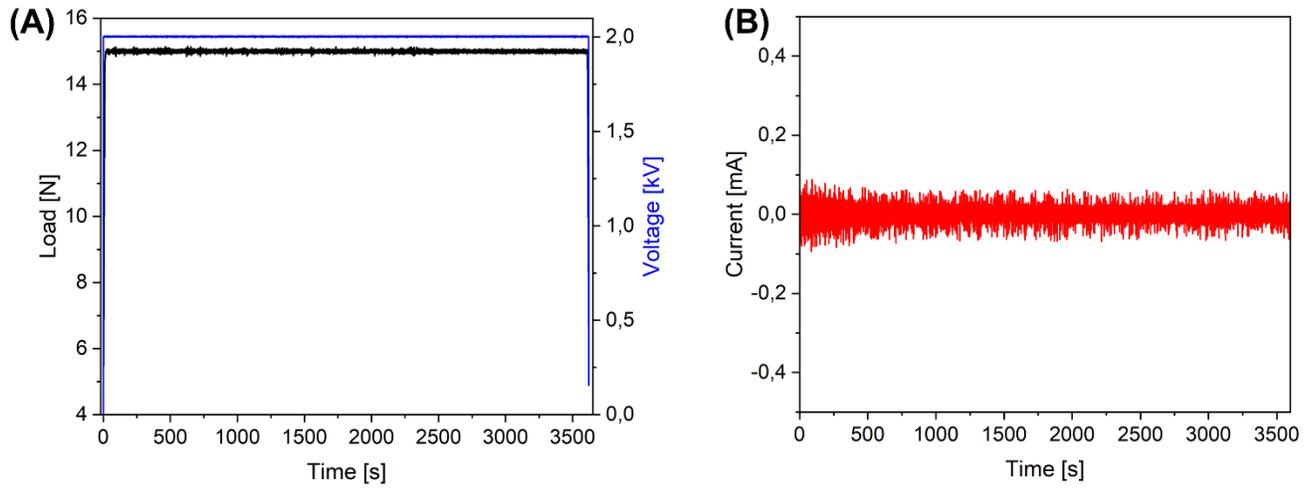

**Figure S5.** (A) Applied load and voltage during electrostatic indentation with 1 hour of load holding. (B) Current through the sample measured during electrostatic indentation with 1 hour of load holding.